\documentclass{article}%

\usepackage{amsmath}%
\usepackage{amsfonts}%
\usepackage{amssymb}%
\usepackage{graphicx}

\begin{document}

\title{{\bf Contribution Prospectives IN2P3 2020}
 {\bf Predictive Physics of Inflation and Grand Unification for and from 
the CMB observations}}
\author{{\bf Principal Author:} 
\\Norma G. SANCHEZ 
\\CNRS LERMA OP-PSL SU, Paris, France 
\\Norma.Sanchez@obspm.fr 
\\https://lerma.obspm.fr/$\sim$sanchez}
\date{\today}
\maketitle

White Paper for the CNRS IN2P3  Prospectives 2020,  GT05 Steering Group 
"Physics of Inflation and Dark Energy", presented at the GT05 IN2P3 Topical Meeting held 
on 9-10 December 2019 in Grenoble, France. 

\begin{abstract}
This White Paper for the CNRS IN2P3 Prospective 2020 \footnote{IN2P3: The French National Institute of Nuclear Physics and Particle Physics of the CNRS. CNRS: The French National Center of Scientific Research} focuses on realistic and timely situations of inflation in connection with the CMB, gravitational and particle physics, adding inter-disciplinarity and unification values within a strongly predictive physical approach. The formulation of inflation in the Ginsburg-Landau approach developed by de Vega and Sanchez [1] and by Boyanovsky, de Vega, Sanchez and Destri [2-4] clarifies and places inflation in the setting of the effective field theories of particle physics. In addition, it sets up a clean way to directly confront the inflationary predictions with the available and forthcoming CMB data and select a definitive model. All CMB + LSS data until now show how powerful is the Ginsburg-Landau effective theory of inflation in predicting observables in agreement with observations, including the inflation energy scale and the inflaton potential, and which has much more to provide in the future. It paves the way to discoveries, new learning and understanding.
\end{abstract}

\section{Summary}

Inflation is today part of the Standard Model of the Universe supported by the cosmic microwave background (CMB), large scale structure (LSS) and other precision cosmological data. It solves the horizon and flatness problems and naturally generates the density fluctuations that seed LSS, CMB anisotropies, and tensor perturbations ({\it primordial gravitational waves of quantum origin:  the primordial gravitons}). Inflation is based on a scalar field $\phi$ (the inflaton) whose potential is fairly flat leading to a slow-roll evolution. 

\bigskip

This white paper to the IN2P3 prospective focuses on the following, realistic and timely situations of inflation in connection with the CMB, gravitational and particle physics, adding inter-disciplinarity and unification values within a strongly predictive physical approach. The formulation of inflation in the Ginsburg-Landau approach [1-3] clarifies and places inflation in the setting of the effective field theories of particle physics. In addition, it sets up a clean way to directly confront the inflationary predictions with the available and forthcoming CMB data and select {\it a definitive model}. 

\bigskip

In the effective theory of inflation {\it`a la Ginsburg-Landau} the inflaton potential is a polynomial in the field $\phi$, highlighting and capturing the relevant scales of the problem  with a {\it universal} form  $V (\phi) = N M^4 w(\phi / [\sqrt{N} M_{Pl}])$, where $M << M_{Pl}$ is the scale of inflation, $N \approx 60 $ is the number of Inflation efolds and  $w$ is dimensionless,  ($M_{Pl}$ is the Planck scale). The slow-roll expansion thus becomes a systematic $1/N$ expansion and the inflaton couplings become {\it naturally small} as powers of the ratio $(M/M_{Pl})^2$. The spectral index $n_s$ and the ratio $r$ of tensor to scalar fluctuations are  $(n_s - 1) = O(1/N)$, $r = O(1/N)$ while the running index turns to be smaller: $dn_s/d (ln k) = O(1/N^2)$. The energy scale of inflation turns out to be $M \simeq 0.7 \;$x$ \;10^{16} GeV$, and a detection of $r$ will allow an important more precise and direct confirmation [10], [14], [15].  A complete analytic study plus the available CMB+LSS data analysis  with fourth degree trinomial potentials shows [3]-[5] :

\begin{itemize}
\item{{\bf(a)} the spontaneous breaking of the   $\phi \rightarrow - \phi$ symmetry of the inflaton potential.} 
\item{{\bf(b)} a {\it lower bound} for $r$ in new inflation: $r \geq 0.023 \; (95 \% CL)$ and $r \geq 0.046 \; (68 \% CL)$.} 
\item{{\bf(c)} The preferred inflation potential is a double well, even function of the field with a moderate quartic coupling yielding as most probable values: $n_s = 0.964, r = 0.041$. This value for $r$ is within reach of forthcoming CMB observations. All data till now clearly prefer new inflation.  Higher order terms in the potential are negligible. In the $(n_s, r)$ diagram this selects the {\it universal banana shaped region} (Destri, de Vega and Sanchez, Phys Rev D77, 043509 (2008);  Annals of Phys 326,  578-603 (2011) bounded by the binomial potential curves.} 
\end{itemize}

Slow-roll inflation is generically preceded by a short fast-roll stage. {\it Fast-roll plus slow roll fit the TT, the TE and the EE mode spectra remarkably well including the low multipoles and the quadrupole suppression} [3],[4]. Moreover, quantum loop corrections reveal very small and controlled by powers of $(H/M_{Pl})^2 \simeq 10^{-9}$, which validates the reliability of the effective theory of inflation [3].  

\bigskip

All  CMB  + LSS data until now [9] show how powerful is the Ginsburg-Landau effective theory of inflation in predicting observables in agreement with observations and which has much more to provide in the future [10],[14],[15]. It paves the way to discoveries, new learning and understanding.

\section{Introduction}

{\bf I.1} The set of robust cosmological data (cosmic microwave background, large scale structures,deep galaxy surveys and other data) support {\bf the Standard (concordance) Model of the Universe} and place Inflation and dark energy (and quasi-de Sitter) stages as pivots of it. Moreover, the physical classical, semi classical and quantum planckian and trans-planckian de Sitter regimes are central to gravitation, quantum physics and particle physics unification. 

\bigskip

Inflation predicts fairly generic features: a nearly gaussian, nearly scale invariant spectrum of (mostly) adiabatic scalar and tensor primordial fluctuations, making the inflationary paradigm fairly robust to the highly precise wealth of data provided by the CMB and other data . 

\bigskip

While more complicated scenarios can be proposed, {\it the CMB data  till now validate the simpler slow roll scenario and points the direction for a possible B-mode detection} which could be at reach. Moreover, future CMB and LSS observations require precise theoretical predictions from inflation, and a deeper understanding of how inflation begins and ends and merging with gravitational physics at higher energies: {\it pre-inflation, trans-planckian physics and their influence on the today structures and cosmological vacuum }[6].

\bigskip

{\bf I.2} Inflation is based on a scalar field (the inflaton) whose homogeneous expectation value drives the dynamics of the expanding cosmological scale factor, plus small quantum fluctuations In single field slow roll inflation, the inflaton potential is fairly flat and it dominates the universe energy during inflation. This flatness leads to a slowly varying Hubble parameter (slow-roll) ensuring a sufficient number of inflation e-folds to explain the homogeneity, isotropy and flatness of the universe, and also explains the {\it almost} gaussianity of the fluctuations as well as the {\it almost}  scale invariance of their power spectrum. 

\bigskip

{\bf I.3} Inflation {\bf generically starts by a short fast-roll stage} where the kinetic and potential energy of the inflaton are of the same order. This is {\bf followed by the slow-roll stage} where the kinetic energy is much smaller than the potential energy. The slow-roll regime of inflation is an attractor of the dynamics during which the Universe is dominated by vacuum energy. Inflation ends when again the  inflaton kinetic energy becomes large as the field is rolling near the minimum of the potential. Eventually, the energy stored in the homogeneous inflaton is transferred explosively into the production of particles via spinodal or parametric instabilities. More precisely, non-linear phenomena eventually shut-off the instabilities and stop inflation. All these processes lead to the transition to the standard radiation dominated era. This is the physical picture of the transition from inflation to the following phase of radiation cosmology.

\bigskip

{\bf I.4 \; Novel phenomena} emerges at the quantum level as a consequence of the lack of kinematic thresholds, among them the phenomenon of inflaton decay into its own quanta. The quantum corrections to the power spectrum are expressed in terms of the observable set $(n_s, r$ and $dn_s/d ln k)$. Trace anomalies dominate the quantum corrections to the primordial power spectrum in a definite direction: They enhance the scalar curvature spectrum and reduce the tensor fluctuations, but they are screened by the overall small factor $(H/M_{Pl})^2$. The quantum loop corrections are very small and controlled by the ratio $(H/M_{Pl})^2$, a conclusion that validates the reliability of the effective field theory approach to inflation.

\section {Effective Field Theory of Inflation: The Ginsburg-Landau approach and its powerful predictions}                

The formulation of inflation in the Ginsburg-Landau approach clarifies and places inflation within the understanding of the effective field theories in particle physics. In addition, it sets up a clean way to directly confront the inflationary predictions with the available and forthcoming CMB data {\it and select a definitive model} [1]-[5].

\bigskip

The theory of the second order phase transitions: the Ginsburg-Landau effective theory of superconductivity, the current-current Fermi theory of weak interactions, the sigma model of pions, nucleons (as skyrmions) and photons, are all successful effective field theories. The  effective theory of inflation is powerful and successful in predicting observable quantities that are contrasted or will be soon contrasted with experiments. 

\bigskip

In the Ginsburg-Landau framework, the potential {\it is a polynomial in the field} starting by a constant term. Linear terms can always be eliminated by a constant shift of the inflaton field. The quadratic term can have a positive or a negative sign corresponding to chaotic or new inflation, respectively. This effective Ginsburg-Landau field theory is characterized by only two energy scales: the scale of inflation $M$ and the Planck scale $M_{Pl} >> M$. In this context we propose a {\it universal} form for the inflaton potential in slow-roll models which encodes the essential physics of the problem is: $V (\phi) = N M^4 w(\chi)$  where $N$ is the known number of e-folds since the cosmologically relevant modes exit the horizon till the end of inflation and $\chi$ is a dimensionless, slowly varying field $\chi  = \phi / (\sqrt {N} M_{Pl})$. The slow-roll expansion becomes in this way a explicit and {\it systematic} $1/N$ expansion. The couplings in the inflaton Lagrangian become {\it naturally} small due to suppression factors arising  as the ratio of the two relevant energy scales in inflation:  The inflation energy scale $M$ and the Planck scale $M_{Pl}$.  

\bigskip

{\bf The whole set of CMB observables and the GUT scale:} We derived in this approach explicit analytic formulae and study in detail the spectral index $n_s$ and the amplitude of the adiabatic fluctuations, the ratio $r$ of tensor to scalar fluctuations, the running index $dn_s/d \;(ln k)$, and  the non-gaussianity  parameter $f_{NL}$, [1],[2],[3]. We use these analytic formulas as hard constraints on $n_s$ and $r$ in the MCMC data analysis. 

\bigskip

The spectral index, the ratio  $r$ of tensor/scalar fluctuations, the running index  $dn_s/d(ln k)$, the amplitude of the adiabatic fluctuations $(n_s - 1) = O (1/N)$, $r = O (1/N)$, $dn_s /d(ln k) =  O (1/N^2)$ , $|\Delta ^R_ {k,\; ad}| \simeq N (M /M_{Pl})^2 $.

\bigskip

Hence, the energy scale of inflation $M$ relates to the amplitude of the scalar adiabatic fluctuations $|\Delta ^R_ {k,\; ad}|$ and using the CMB data for it, we find $M \simeq 10^{16} \;GeV$. A further and precise determination of $M$ will come also from the detection of $r$, (B modes) directly related to $V (\phi)$.

\bigskip

Therefore, {\it the microscopic theory of inflation is expected to be a GUT in a cosmological space-time}. The relation between the effective inflation theory and the microscopic fundamental GUT is akin to the relation between the effective Ginsburg-Landau superconductivity  theory and the microscopic BCS (Bardeen-Cooper-Schrieffer) theory, or like the relation of the O(4) sigma model as effective theory of pions, photons and chiral condensates with quantum chromodynamics (QCD). {\it Whatever the microscopic model for the early universe (GUT theory) would be, it should include inflation with the generic features we know today successfully tested by observations.}

\bigskip

{\bf The Inflaton potential from the CMB data:} The potential which best fits the present data for red tilted spectrum $(n_s < 1)$ and which best fit the data (a small $r < 0.08$) is given by the trinomial potential with a negative $\phi^2$ term, that is new inflation that is, symmetry breaking with negative concavity $V"(\phi) < 0$. In new inflation we have the upper bound $r \leq 0.08$, and the {\it lower bound} $r > 0.023$ . More precise measurements of $n_s$ together with better data on $r$ and $dn_s/d (ln k)$ will permit to better select the best inflation model. This will allow to improve the prediction that a broken symmetric inflaton potential with moderate nonlinearity (new inflation) best describes the data {\it and select the definitive model}.

\bigskip

The MCMC analysis of the best CMB+LSS data with the Ginsburg-Landau effective theory of inflation showed [3],[5]: {\bf (i)} The data strongly indicate the breaking (whether spontaneous or explicit) of the  $\phi \rightarrow - \phi $ symmetry of the inflaton potentials both for new and for chaotic inflation. {\bf (ii)} Trinomial new inflation naturally satisfies this requirement and provides an excellent fit to the data. {\bf (iii)} Trinomial chaotic inflation only produces the best fit in a very narrow corner of the parameter space. {\bf (iv)} {\it The chaotic potential is almost certainly ruled out (at more than $95 \; \% CL$)}. {\bf(v)} The above results and further physical analysis {\it we conclude that new inflation gives the best description of the data.} [ref 3 and our refs therein].
{\bf (vi)} We find a {\it lower bound} for $r$ within trinomial new inflation potentials: $r > 0.023 \;(95 \% CL)$ and $r > 0.046 \;(68 \% CL)$. {\bf(vii)} {\it The preferred new inflation potential is a double well, even function of the inflaton field with a moderate quartic coupling}. This new inflation model yields as most probable values: $n_s = 0.964$, $r = 0.051$. This value for $r$ is within reach of forthcoming CMB observations. {\it The inflaton field exits the horizon in the negative concavity region   $V"(\chi) < 0 $ intrinsic to new inflation}. We find for the best fit, $M = 0.543 \; $x$ \; 10^{16}$ GeV for the scale of inflation and $m = 1.21 \;$x$\; 10^{13}$ GeV for the inflaton mass.  

\bigskip

• Higher degree $2n$ terms ($n > 2$) in the inflaton potential do not affect the fit in a significant way in new inflation. The window of agreement with the data narrows for growing n in chaotic inflation.  {\it All members of the new inflation family predict a small and negative running}
$$-4 (n + 1) \; 10^{-4} \leq  dn_s / d(ln k) \; \leq \; -2 \; 10^{-4}$$

{\bf Further Physical Implications from the CMB data: Energy scale of Inflation and the mass of the inflaton:} The Ginsburg- Landau (polynomial) realization of  the inflaton potential fits the amplitude of the CMB anisotropy remarkably well and reveals that the Hubble parameter, inflaton mass and non-linear couplings are of {\it the see-saw form in terms of the small ratio $M / M_{Pl}$}. 

\bigskip

It appears clearly that it is highly unnatural to consider only monomial inflaton potentials, and to drop the quadratic term $\phi^{2}$, since this would be to exactly choose $m^2 = 0$. In fact, the CMB data  unfavor the monomial $\phi^4$ potential and support a polynomial inflaton potential, a binomial potential being enough for the present data. Excluding the quadratic mass term in the potential $V (\phi)$ implies to fine tune to zero the mass term of the inflaton, only justified at isolated (critical) points. 

\bigskip

Therefore, from the physical point of view, the pure monomial potentials $\phi^{2n}$ is a weird choice. {\it The fact that the pure $\phi^4$ potential is clearly disfavored by the CMB data implies a lower bound on the inflaton mass $m  \geq  10^{13} \; GeV $}. Therefore, updating knowledge and understanding, we conclude that in the analysis of inflation with high precision CMB data: {\it using monomial inflation potentials (whatever its degree) is a poor choice}. The best choice from  theory and data is the binomial or trinomial potential with negative $\phi^2$ term (new inflation).

\bigskip

The mass of the inflaton $ m \simeq 10^{13} \; GeV $ can be related with the scale mass of Grand Unification $M_{GUT}$  by a see-saw type relation, $ m =  M_{GUT}^2 / M_{Pl} $. The massless fields alone cannot describe inflation  with the observed amplitude of the CMB anisotropies. 

\bigskip

{\bf Smallness of the inflaton couplings:}   The quartic coupling is $\lambda = (M /M_{Pl})^4/N $. {\it The smallness of the non-linear couplings is not a result of fine tuning but a natural consequence of the validity of the effective field theory of inflation} [1],[2],[3].
The quantum expansion in loops is therefore a double expansion on $(H/M_{Pl})^2$ and $1/N$.  {\it The form of the potential which fits the CMB+LSS data and is consistent with slow-roll implies the small values for the inflaton self-couplings}. 

\bigskip

{\bf Departures from scale scale invariance:} In Inflation, the near departure of scale invariance of the fluctuations introduces a natural regularization to the strong infrared behavior of de Sitter space-time through the difference between  the slow roll parameters expressed as $ \Delta = (n_s - 1)/2 + r/8 $. [3]

\bigskip

{\bf The running of the spectral index, its value and sign: $dn_s/d (ln k)$:}  All members of the 
new inflation family predict a small negative running:
$ - 4 (n + 1) \; $x$ \;10^{-4} \; \leq \; dn_s/d (ln k) \; \leq \; - 2 \; $x$ \; 10^{-4} $. Because the range of the cosmologically relevant modes is $\Delta (ln\; k) < 9$, we get $\Delta n_s < 9/N^2 \simeq 0.0025$. 
Therefore, the effective theory of slow-roll inflation indicates that the detection of 
the running calls for measurements of $n_s$ with a one per thousand precision on a wide range of wave numbers.

\bigskip

{\bf Smallness of Non-gaussianity:} Non-gaussianity is predicted of the order  $f_{NL} \simeq (1 / N) \sim 0.02$, in the effective single-field slow-roll inflation. Self-interactions of the fluctuations of the scalar field in turn lead to non-gaussianities which are characterized by a non-vanishing bi-spectrum. The connection between the self-decay of inflaton fluctuations and the bi-spectrum in single field slow roll inflation was established in ref [3].  {\it CMB data $ f_{NL} < 6 $ (Planck 2018, 2019) [9] confirm our predictions from WMAP 2003 that  $f_{NL}$ is very small. Our conclusion, comforted by all available observational data is that our predictive theory points towards a very small amount of primordial non-gaussianity $f_{NL} \simeq 0.02$}. 

\bigskip

{\bf Tensor/scalar ratio $r$ by the forthcoming CMB experiments and Forecast for detection}. The detection of $r$  would be the first detection of primordial gravitational waves [10], [14,15]. In addition, since such primordial gravitational waves were born as quantum fluctuations, this would be the first detection of gravitons, namely, quantized gravitational waves at tree level. Such detection of the primordial gravitational waves will test our prediction $ r = 0.04 - 0.05 $ [1]-[5],  based on the effective theory of slow-roll inflation [1]-[5] (broken symmetric binomial and trinomial potentials and the set of their observable implications).

\bigskip

{\bf Effects of Generic Initial conditions:} We find and compute the transfer function $D(k)$ which encodes the effect of generic initial conditions on the power spectra  [3], [4]. (Usually, scalar (curvature) and tensor perturbations are studied only with asymptotic vacuum initial conditions {\it within} the slow-roll stage). {\it The observable effects from initial conditions are more prominent in the low CMB multipoles}. The effects on high $\ell$-multipoles are suppressed by a factor $\sim 1/\ell ^ 2$ due to the large $k$ fall-off of $D(k)$. Hence, a change in the initial conditions for the fluctuations during slow roll can account for the low observed value of the CMB quadrupole.

\bigskip

{\bf Physical relevance of the low $\ell$  part  of the TT, TE and EE spectra:} Although there are no statistically significant departures from the slow roll inflationary scenario at small angular scales, the CMB data confirm again the low quadrupoles and suggest that it cannot be completely explained by galactic foreground contamination. The low value of the quadrupole has been an intriguing feature on large angular scales since first observed by COBE/DMR, confirmed by WMAP and  by the following CMB data till now. The low $\ell$ part of the TT, TE and EE together become a {\bf crucial test and  source of information} for the initial conditions and fast-roll of inflation, which next CMB data and missions [10], [14], [15] should be able to provide. 

\bigskip

{\bf Testing the Fast-roll Inflation stage preceding Slow-roll and the low  CMB multipoles:} Slow-roll inflation is {\it generically} preceded by a short fast-roll stage [3],[4]. The vacuum initial conditions usually used for slow roll fluctuations are in fact the natural initial conditions for {\it the fast-roll} fluctuations.  The physical reason of quadrupole suppression is the following: During fast-roll, the potential in the wave perturbation equations is {\it purely attractive} and leads to a {\it suppression} of the curvature and tensor CMB quadrupole s with respect to the slow-roll stage in which it is purely repulsive.   

\bigskip

The CMB + LSS data analysis {\it including the fast-roll inflationary stage} shows that the quadrupole mode exits the horizon about $0.2$ efold before fast-roll ends and therefore its amplitude gets suppressed. In addition, fast-roll fixes the initial inflation redshift to be $z_{init} = 0.9\; $x$\; 10^{56}$ and the total number of e-folds of inflation to be $N_{tot} \simeq 64$. {\it Fast-roll plus slow roll fit the TT, the TE and the EE mode spectra remarkably well including the low multipoles and the quadrupole suppression}.    
                                                                                                                         \bigskip
																																																												
{\bf Quantum corrections to Inflation:} A thorough study [3]  of the quantum loop corrections reveals that they are very small and controlled by powers of $(H/M_{Pl})^2 \simeq 10^{-9}$ , where $H$ is the Hubble parameter during inflation, a conclusion that validates the tree level results and the reliability of the effective field theory approach to inflation. Quantum corrections to the power spectra are expressed in terms of the CMB observables: $n_s, r $ and $dn_s/d (ln k)$. {\it Trace anomalies, especially the graviton part}, dominate the  quantum corrections to the primordial spectra in a definite direction: they enhance the scalar curvature fluctuations and reduce the tensor fluctuations.  They are {\it all  screened} by the overall factor $(H/M_{Pl})^2$ and higher orders are still more screened. {\it The re-summation and full quantum spectra} containing the semiclassical inflationary spectra as a particular case,  have been recently obtained in a more general framework and confirm these results [6]. 

\bigskip

{\bf Grand Unification energy scale appears in three important physical situations:} {\bf(a)} The scale of Grand Unification of strong and electroweak interactions.  {\bf (b)} The  large energy scale in the see-saw formula for neutrino masses to explain neutrino oscillations. {\bf (c)} The energy scale of inflation, $M \simeq 10^{16}\; GeV$.  This coincidence suggests a physical link between the three areas. Moreover, in the standard model of electromagnetic, weak and strong interactions, the renormalization group yields that the unification of the three couplings is better reached in supersymmetric extensions of the standard model. The inflaton potential suggests that the supersymmetry  breaking scale $m_{susy}$ turns out to be at the GUT scale:  $m_{susy} \approx M_{GUT}$ and  best CMB data and polarization should improve and allow go deeper into this implication. 

\bigskip
 
{\bf Before Inflation and Beyond Grand Unification: Pre- Inflation, planckian and trans-planckian physics:} Within a prospective at 2020- 2030, it is important to keep in mind that in the primordial phases of the Universe, besides inflation and its GUT energy scale at time scales of $10^{-32}$ sec, there is room for {\it higher energy scales at earlier times} which are of the order of the Planck fundamental scale $10^{19} \; GeV$ at $10^{-44}$ sec and beyond, i.e., {\it the so-called trans-planckian regime}. This quantum phase and its late imprints is a targeted field of study in quantum unification theories, gravitation and cosmology. The understanding of dark energy within the standard concordance model description, (namely the cosmological vacuum energy, cosmological constant problem, or dynamical dark energies for instance), is at the center of these studies. {\it The past remote states of the universe before inflation are the natural setting for planckian and trans-planckian energies} [6].

\bigskip

{\bf Inflation and Dark Energy:} The most direct and simplest explanation of the origin of dark energy which is compatible with all the available data is the cosmological constant, the so called "cosmological constant problem": The theoretical vacuum energy value estimated from quantum particle physics is about $10^{122}$ times the observed value today. Such a huge difference between the two values could physically correspond to two different vacuum states of the universe, namely at two different epochs of its evolution: One being the classical very large universe today, the other being the very early quantum cosmological vacuum [6]. The low value of Lambda or vacuum energy density today corresponds to a classical large scale low energy diluted universe essentially dominated by voids and super-voids as the set of large scale observations concordantly and independently shows. On the other hand, the high quantum estimated value of Lambda could correspond to the high energy and highly dense small scale very early quantum vacuum [6]. This links with the {\it Action Nationale Dark Energy} [12] and ESA voyage 2050  [11],[12].

\bigskip

{\bf Conclusions and further questions:} This short paper presents the state of the art of the effective theory of inflation and its successful confrontation with the CMB and LSS data.  It paves the way for a better determination of the inflaton potential, GUT implications, primordial gravitons amount constraints or its detection, and completes the generic and robust predictions of slow-roll and fast-roll inflation beyond present knowledge. Forthcoming observations of CMB anisotropies and polarizations as well as large scale surveys with ever greater precision and extent will provide a substantial body of high precision observational data. The robust physical approach presented here is well prepared and timely to allow to extract the best scientific benefit and interpretation from such data. Studying the observational consequences of the classical and quantum phenomena presented in this prospective IN2P3 paper will therefore prove a worthwhile endeavor.

\bigskip

{\bf Outlook  and Future Perspectives:}  We can highlight as perspectives for a foreseeable future: The CMB data till now show how powerful and successful is the Ginsburg-Landau effective theory of inflation in predicting observables that are contrasted successfully to observations and pave the way to fruitfully extract high benefit from with a strategy of discoveries, namely the B-mode detection, the probe of Grand Unification scale and the hint of supersymmetry breaking. Neutrino oscillations offer an example of macroscopic quantum coherence and keV sterile neutrinos a dark matter candidate to solve the galactic scale problems. The Novel quantum phase before inflation may hold clues to the problem of dark energy.  This interdisciplinarity merges together modern cosmology, quantum physics, particle physics, neutrino oscillations and keV sterile neutrinos as dark matter candidates.

\bigskip

{\it Is also very important and timely to investigate the pre-Inflation phases and its higher energies to go beyond present knowledge} [6],[7],[8].

 \section{References}

1. Single Field Inflation models allowed and ruled out by the WMAP data, \\
H. J. de Vega and N. G. Sanchez, Phys. Rev. D74, 063519 (2006) \\
https://inspirehep.net/record/713964 
 
\bigskip

2.  Clarifying Inflation Models: Slow-roll as an expansion in 1/Nefolds, \\
D. Boyanovsky, H. J. de Vega, and N. G. Sanchez, Phys. Rev. D73, 023008 (2006). 
https://inspirehep.net/record/688214 
	
\bigskip
	
3.  The Effective Theory of Inflation in the Standard Model of the Universe and the CMB+LSS data analysis, 
D. Boyanovsky, C. Destri, H. J. de Vega, N. G. Sanchez, Int.  J. Mod. Phys. A24, 3669 (2009)\\
https://inspirehep.net/record/810323   

\bigskip 
       
Quantum corrections to the inflaton potential and the power spectra from superhorizon modes and trace anomalies, 
D. Boyanovsky, H. J. de Vega, N. G. Sanchez, Phys.Rev. D72, 103006 (2005) \\
https://inspirehep.net/record/688215 

\bigskip

Quantum corrections to slow roll inflation and new scaling of superhorizon fluctuations,                          
D. Boyanovsky, H. J. de Vega, N. G. Sanchez, Nucl.Phys. B747, 25-54 (2006).                                            https://inspirehep.net/record/679335 

\bigskip

4.  The pre-inflationary and inflationary fast-roll eras and their signatures in the low CMB multipoles,  
C. Destri, H. J. de Vega , N. G. Sanchez, Phys.Rev. D81, 063520 (2010).
https://inspirehep.net/record/840213    

\bigskip

The CMB Quadrupole depression produced by early fast-roll inflation: MCMC analysis of WMAP and SDSS data, 
C. Destri, H. J. de Vega , N. G. Sanchez, Phys.Rev.D78:023013,(2008).                                               
https://inspirehep.net/record/783425 

\bigskip

5.  Forecast for the Planck precision on the tensor to scalar ratio and other cosmological parameters,           
C. Burigana, C. Destri, H. J. de Vega, A. Gruppuso, N. Mandolesi, P. Natoli, N.G. Sanchez, 
The Astrophysical Journal  724, 588 (2010).
https://inspirehep.net/record/850564 

\bigskip

6.  New quantum phase of the Universe before inflation and its cosmological and dark energy implications, 
N. G. Sanchez, Int. J. Mod. Phys. A34, 1950155 (2019) \\
https://inspirehep.net/record/1756910

\bigskip

7.  New Quantum Structure of the Space Time,                                                         
N. G. Sanchez, Gravitation and Cosmology 25, Issue 2, pp 91–102 (2019) (Springer). \\
http://inspirehep.net/record/1739920 

\bigskip

8. The classical-quantum duality of nature including gravity,
N. G. Sanchez, Int. J. Mod. Phys. D28, 1950055 (2019)\\
https://inspirehep.net/record/1662162 

\bigskip

9. Planck Collaboration: arXiv:1807.06211 (2018), arXiv:1905.05697 (2019).

\bigskip

10. LiteBIRD-Europe http://www.litebird-europe.eu/      \\
LiteBIRD-France  https://indico.in2p3.fr/event/19378/overview \\
arXiv2001.01724 (2020) 

\bigskip

11. Voyage 2050 White Paper Microwave Spectro-Polarimetry of Matter and Radiation across Space and Time,  
J. Delabrouille et al  (July 2019). \\
https://www.cosmos.esa.int/web/voyage-2050/white-papers 

\bigskip

12. Voyage 2050 White Paper  Gravitation And the Universe from large Scale-Structures. The GAUSS mission concept .Mapping the cosmic web up to the reionization era,   A. Blanchard et al (July 2019). \\
https://www.cosmos.esa.int/web/voyage-2050/white-papers 

\bigskip

13. Euclid http://sci.esa.int/euclid/; \;  DESI: http://desi.lbl.gov ;\;                                   
WFIRST: https://wfirst.gsfc.nasa.gov 

\bigskip

14. CMB-S4 Decadal Survey APC White Paper, (July 2019), 
Bull. Am. Astron.Soc. 51 (2019) no.7, 209; 
http://inspirehep.net/record/1748019  

\bigskip

15.  CMB Appec programme (APC Paris, Sept 2019) \\ 
https://indico.in2p3.fr/event/19414/  

\bigskip

16. Simons Observatory:  https://simonsobservatory.org/ ;  	JCAP 1902 (2019) 056;
https://inspirehep.net/record/1689432  

\bigskip

17. LISA Mission: https://www.lisamission.org/; http://sci.esa.int/lisa/; https://lisa.nasa.gov/; LIGO and Virgo Collaborations: B.P. Abbott et al., Phys. Rev. Lett. 116, 061102 (2016); 
DES and LIGO/Virgo Collaborations: M.Soares-Santos, A.Palmese et al,
The Astrophysical Journal Letters, Volume 876, 1, L7 (2019);
https://inspirehep.net/record/1712453

\bigskip

\section {Acknowledgements}
This work was performed in LERMA-CNRS-Observatoire de Paris- PSL University-Sorbonne University. 
The author acknowledges the French National Center of Scientific Research (CNRS) for Emeritus Director of Research contract, the IN2P3 for the timely and appropriate Prospective 2020 "Physics of Inflation and Dark Energy" of the Steering Group GT05, and the organizers of the Topical GT05 Meeting in Grenoble which arised so interesting discussions and new perspectives in connection with the CMB, large scale structures, $H_0$ and related current topics.

\end{document}